  \documentstyle [11pt]{article}
  \newcommand{\be}{\begin{equation}}
  \newcommand{\ee}{\end{equation}}
  \newcommand{\bea}{\begin{eqnarray}}
  \newcommand{\eea}{\end{eqnarray}}
  
\begin{document}
  \title{
  Front Structures in a Real Ginzburg-Landau
  Equation Coupled to a Mean Field}
  \author{Henar Herrero \\
  \small Departamento de F\'{\i}sica y Matem\'atica Aplicada,
  Facultad de Ciencias,\\
  \small Universidad de Navarra, 31080 Pamplona, Navarra, Spain\\
   Hermann Riecke \\
  \small Department of Engineering Sciences and Applied Mathematics \\
  \small Northwestern University, Evanston, IL 60208, USA}

  \maketitle
  \baselineskip=15pt

  \begin{abstract}

  Localized traveling wave trains or pulses have been observed in various
  experiments in binary mixture convection. For strongly negative separation
ratio,
  these pulse structures can
  be described as two interacting fronts of opposite orientation.
  An analytical study of the front
  solutions in a real Ginzburg-Landau equation coupled to a mean field
  is presented here as a first approach to the pulse solution. The additional
mean field
 becomes important when the mass diffusion in the mixture
 is small as is the case in liquids.
Within this framework it can lead to a hysteretic transition between slow and
fast fronts when the Rayleigh number is changed.
  \end{abstract}

\ \\

\centerline{To appear in: J. Bif. and Chaos}
  \newpage
  \bigskip

  \section{\bf Introduction}

  \setlength{\parindent}{15pt}
  \setlength{\parskip}{12pt}

  Convection in binary mixtures occurs through a Hopf bifurcation for
  sufficiently strong coupling between the thermal and the concentration
  field. This bifurcation is usually subcritical and leads to traveling
  waves. Strikingly, they can form localized-wave trains (LTW). Two classes of
LTW seem
  to exist: a discrete set of relatively short, pulselike structures, which
 resemble
 solitons in the nonlinear Schr\"odinger equation, and  longer LTW,
  which resemble a pair of fronts connecting
  the convective and the conductive state. Previously, it seemed that the
latter
  have arbitrary length. Very recent experiments showed, however, that this is
not the case and that their stable length is fixed, too  [Kolodner, 1993].
In both cases the drift
  velocity is very small.  Theoretical efforts to understand the LTW within
  the framework of reduced equations like the complex Ginzburg-Landau equation
  (CGL) have focussed on different points of view:
 1) considering the CGL as a perturbation of the nonlinear
  Schr\"odinger equation, one can show the stability of short LTW
 [Thual \& Fauve, 1988, Fauve \& Thual, 1990];
  2) in the weakly dispersive limit, i.e. when the imaginary parts of all of
the
  coefficients in the CGL are very small, bound pairs of fronts can be stable
with their distance diverging when the imaginary parts go to zero
 [Hakim {\it et al.}, 1990, Hakim \& Pomeau, 1991].
 In both approaches the discrepancy between the calculated linear group
velocity
  $s$ and
 the considerably smaller observed
 velocity $v$ of the LTW requires that nonlinear contributions balance
 the linear group velocity over a range of parameters.

 Taking into account full numerical simulations of the Navier-Stokes
  equations, which show the relevance of a large-scale concentration field
 [Barten {\it et al.}, 1991],
  it has been suggested  [Riecke, 1992] that the smallness of the drift
velocity
 $v$ of the LTW
  is due to an additional slow time scale in the binary-mixture system.
 It arises due to the smallness of the
  Lewis number ${\cal L}$, which measures the ratio of molecular to thermal
diffusion, and leads to an additional dynamical degree of freedom -- beyond the
convective amplitude, which corresponds to a concentration mode.
 Numerical simulations of the resulting coupled PDE show
 that for small ${\cal L}$ the pulse velocity is drastically reduced, and the
effect
  of the group velocity $s$ on the pulse velocity is strongly diminished as
  compared to the conventional CGL.

 In both cases described above  [Fauve \& Thual, 1990,Hakim \& Pomeau, 1990]
 the localization mechanism is
 due to the dispersion of the waves, i.e. no localization would be possible if
the imaginary
 coefficients in the CGL vanished identically.
The strong effect of the concentration mode on the velocity
 of the LTW raises the question whether localization could arise {\it solely}
 due to
the concentration mode without any dispersion. Here we make a first step
towards
  answering this question.
 We investigate analytically single front solutions in the presence of this
  mean field and focus
 on the dispersionless case.

  The equations proposed in  [Riecke, 1992] are
  \begin{eqnarray}
  \partial_T A + s \partial_X A &=& d \partial^2_X A + (a+fC)A + cA \mid A
\mid^ 2
  + p A \mid A \mid^4 \label{eqa}\\
  \partial_T C &=& \delta \partial^2_X C - \alpha C
  + h \partial_X \mid A \mid^2.  \label{eqc}
  \end{eqnarray}
  They represent a minimal model for the mechanism of interest. The complete
equations
up to cubic order, including the values of the coefficients for
free-slip-permeable boundary
 conditions, are given in  [Riecke, 1992]. Without dispersion,
the real Ginzburg-Landau equation
  for the real amplitude, without coupling to the $C$-field and after
  a rescaling can be written as
  \begin{equation}
  \partial_t A + s \partial_x A = \partial^2_x A - \frac{\partial {\cal V}}
{\partial A}
  \end{equation}
   with
  \begin{equation}
  {\cal V}(A)=\frac{1}{2}A^2 - \frac{c}{4}A^4 + \frac{1}{6}A^6.
  \end{equation}
  In the subcritical case $c \ge 0$ the potential ${\cal V}$ has two relative
  minima for $A \geq 0$, one at $A=0$ and another at $A=A^*$. At
  $c=c_c(\equiv \frac{4}{\sqrt{3}})$ the
  potential has the same value for the conductive state $A=0$ and for the
convective state $A=A^*$, and the front
  solution connecting the conductive state to the convective state
is stationary. For
  $c > c_c$  the $A=A^*$ state invades the trivial state. Due to the attractive
 interaction between the fronts, domains of definite size exist, they are,
however, unstable.
  The size of these domains
  diverges when $c$ tends to $c_c$ and in this limit the localized unstable
  domain can be described perturbatively as a bound state of two fronts.

  \section{\bf Front Solutions}
  The front between the trivial state and the convective state for $C=0$ is at
  rest for $c=c_c$ and has the form
  \begin{equation}
  A_L(x-x_0) = A^*\left(\frac{1}{2}(1+\tanh(\frac{x-x_0(t)}{\xi}))\right)^{1/2}
  \end{equation}
  where $A^{*2} =\sqrt{3}$ and $\xi=1$.
  If $c \neq c_c$ and $C \neq 0$ the front starts to move.
 In a frame moving with velocity $v$, in which the front is stationary,
 Eqs.(\ref{eqa})-(\ref{eqc}) read
  \begin{eqnarray}
  (s - v)A' & = & A'' - A + cA^3 - A^5 + CA \label{eqa1}\\
  -vC' & = & \delta C'' - \alpha C + h \partial_x A^2.\label{eqc1}
  \end{eqnarray}
 Here $f$ has been absorbed into $h$ and $C$.
  Assuming $c_1 \equiv c-c_c$ and $C$ small, $A_L$ is a zeroth
  order approximation to the moving front,
 \begin{equation}
 A=A_L+\epsilon A_1 + ... \quad \quad C = \epsilon C_1 + ...
 \end{equation}
 and $s-v=O(\epsilon)$. In order to solve the inhomogeneous equation for $C$ we
 assume strong advection and
 expand $C$ in $v^{-1}$. In this limit Eq.(\ref{eqc1}) becomes a first-order
equation and
only
 the boundary condition on $C$ ahead of the front can be satisfied.
 Thus, in the case $v > 0$ ($v < 0$) one has $C \rightarrow 0$ for
 $x \rightarrow \infty$ ($x \rightarrow -\infty$). In order to satisfy the
other
  boundary
 condition as well one has to take into account that in back of the front,
where
  $\partial_x
 A^2$ goes to zero, the term $\alpha C/v$ becomes dominant. This can be done
using
 a suitable matching procedure. To obtain the velocity of a single front
this is, however, not necessary. Thus, one obtains
 \begin{eqnarray}
  C &=& -\frac{h}{v}A_L^2 + \frac{h \delta}{v^2}(A_L^2)'-
  \frac{h \alpha}{v^2} \int_{-\infty}^{x}A_L^2 dx + O(\frac{1}{v^3}), \qquad v
<
  0,\\
  C &=&-\frac{h}{v}(A_L^2-A^{*2}) + \frac{h \delta}{v^2}(A_L^2)'+
  \frac{h \alpha}{v^2} \int_{x}^{\infty}(A_L^2-A^{*2}) dx + O(\frac{1}{v^3}),
  \qquad v > 0,
  \end{eqnarray}
  This expansion is inserted into the equation for $A$. The leading order
equation is
 solved by $A_L$. At $O(\epsilon)$ a solvability
condition
 arises due to the translational symmetry of the system. Projection onto the
left zero-eigenvector $\partial_xA_L$ yields an equation for the velocity $v$,
 \begin{equation}
 s = v + \sqrt{3}\left(c_1 + \frac{h}{v}\right)+ \frac{2 \sqrt{3} \delta h}{3
v^2}
   -\frac{\sqrt{3} \alpha h}{v^2}.\label{e:sv}\\
 \end{equation}
 Thus, for $h > 0$ the absolute value of the velocity ($|s|<|v|$) of a
 leading front ($v<0$) is increased by the concentration field at leading
order,
 whereas that of a trailing front ($v>0$) is decreased. Considering that
positive values
 of $C$ increase the local growth rate of $A$, this is expected and illustrated
 in fig.1.
 There, two typical numerical solutions of Eqs.(\ref{eqa})-(\ref{eqc})
are shown for $v<0$ and
 for $v>0$. In both cases the positive concentration mode supports the growth
of the convective state.

 Eq.(\ref{e:sv}) suggests a hysteretic transition from fast fronts to slow
fronts when the
 group velocity $s$ is decreased with the latter traveling opposite to the
group
  velocity.
 Although this transition occurs outside of the regime of validity of
Eq.(\ref{e:sv}),
this is in fact found in numerical simulations of Eqs.(\ref{eqa})-(\ref{eqc})
  [Herrero \& Riecke, in preparation] and is
 reminiscent of a similar transition for pulses  [Riecke, 1993].
 Experimentally of interest should be the fact that this transition in the
front velocity can also be achieved by changing $c_1$. Due to our scaling of
Eqs.(\ref{eqa})-(\ref{eqc})
 this is equivalent to changing the Rayleigh number. Thus, the velocity
of a single front is expected to be strongly nonlinear in the Rayleigh number
whereas without the concentration mode it would be linear to lowest order.

 In conclusion, we have investigated the influence of an additional mean field
on the dynamics
 of propagating fronts in a real Ginzburg-Landau equation without
dispersion. This is considered a
 first step
 towards the understanding of localized traveling-wave pulses which have been
found in
 simulations of these equations and which appear to be similar to localized
waves found
 in binary-mixture convection at strongly negative separation ratio. To treat
the interaction
 of such fronts one has to take into account the slow decay of the
concentration
  mode
 behind the front. In a limiting case in which the fronts are very steep this
decay can be
 seen to be the dominant contribution to the interaction and can lead to stable
 bound
 states of two fronts, i.e. stable localized waves
 [Herrero \& Riecke, in preparation].

  {\bf Acknowledgements}

  H.H. was supported by a grant of the F.P.I. program Ref.AP90 09297081
  (M.E.C.). H.R. gratefully acknowledges the hospitality of the University of
Navarra.
  The work of both authors was supported by DOE through grant
DE-FG02-92ER14303.
 We gratefully acknowledge also discussions with W.L.Kath.

\newpage

\section{Figure Captions}

{\bf Figure 1:} Typical front for $c=-2.309$, $\delta = 0.009$,
$\alpha =0.01$ and $h=0.03$.\\
 a) $s=0.15$, b) $s=0.5$.

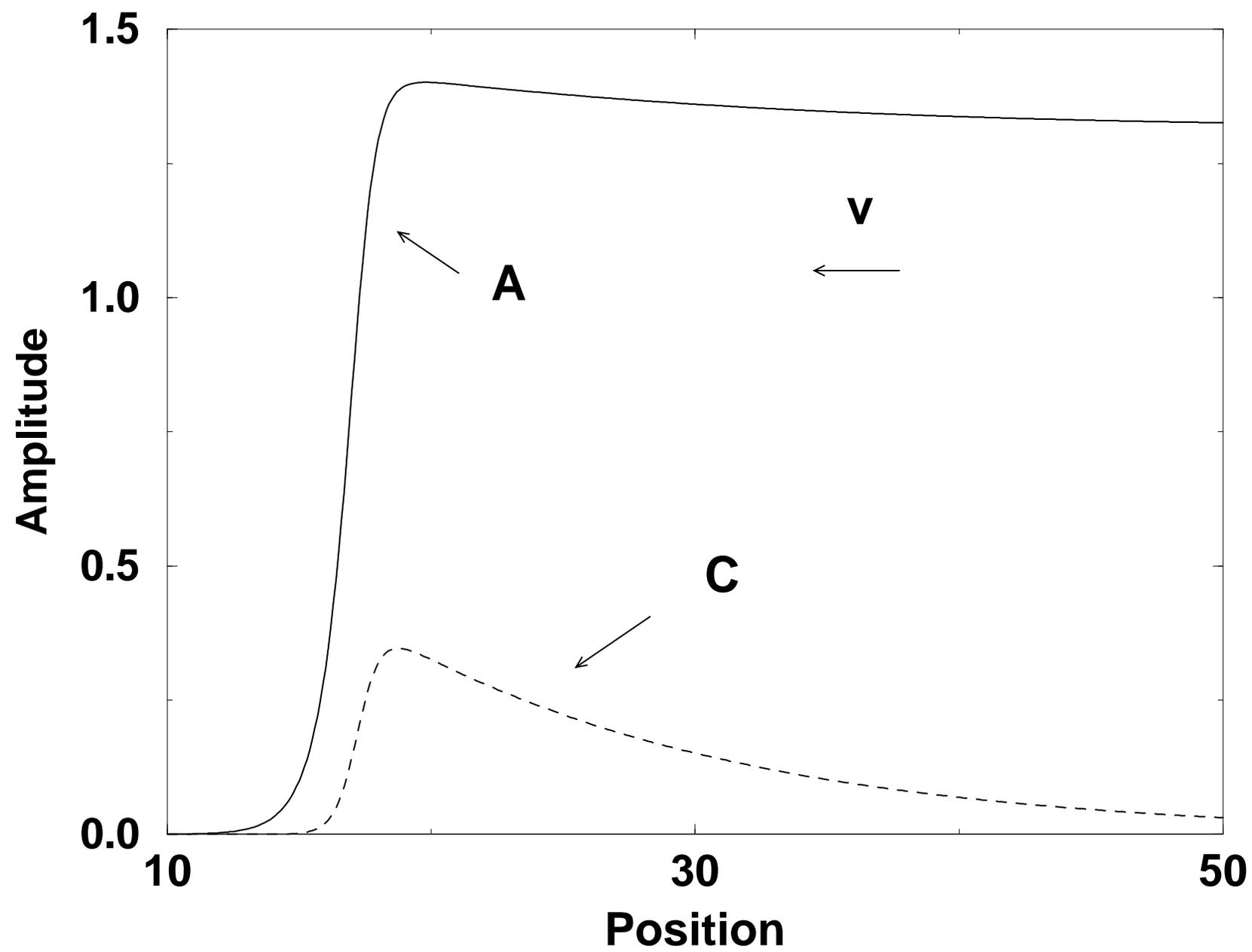
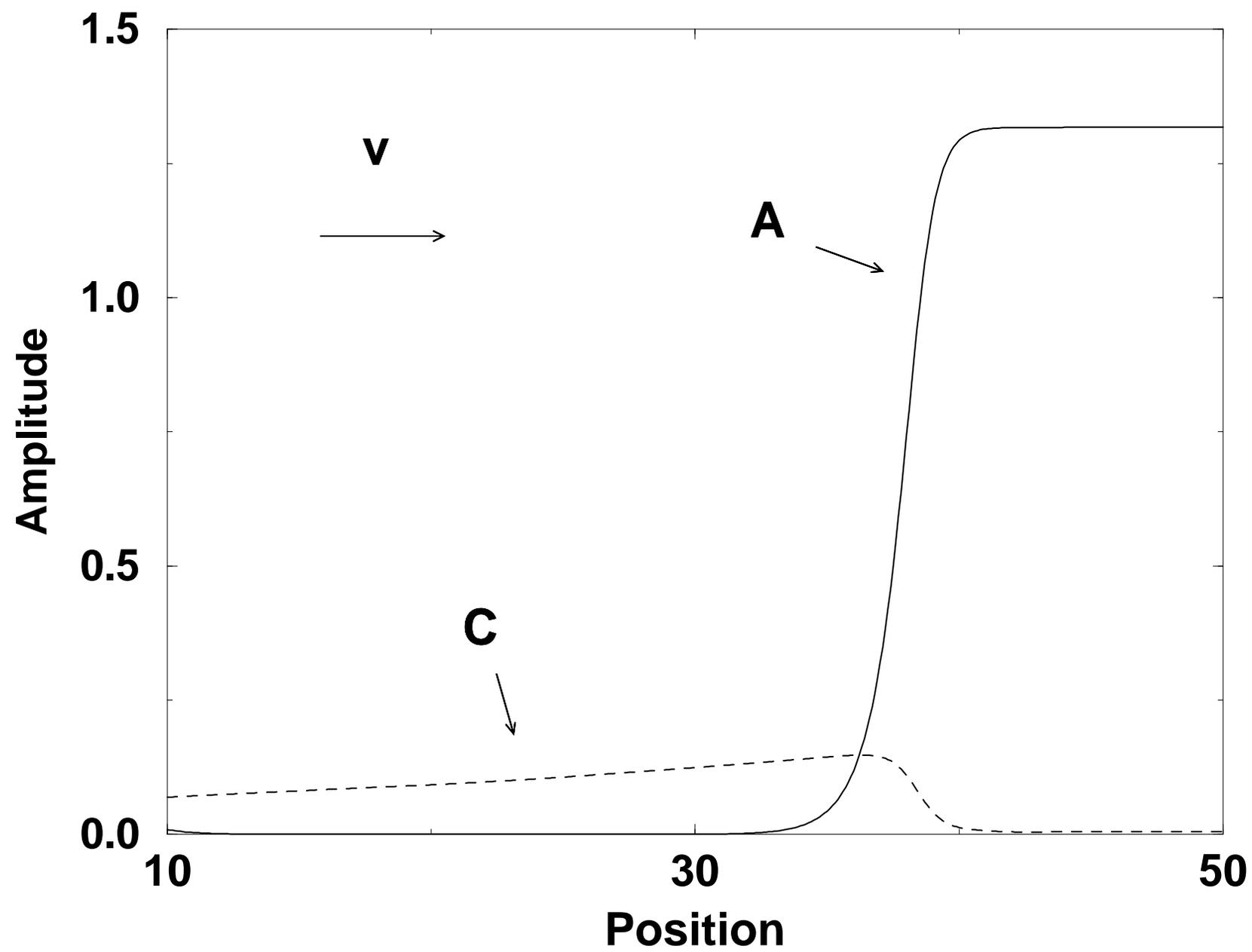
\end{document}